\documentclass[twocolumn,showpacs,preprintnumbers,amsmath,amssymb]{revtex4}  
\begin{document}

\title{Conflict between trajectories and density description: 
the statistical source of disagreement}
\author{Paolo Allegrini$^1$, 
Paolo Grigolini$^{2,3,4}$, Luigi Palatella$^4$, Angelo Rosa$^5$}
\affiliation{$^1$Istituto di Linguistica Computazionale 
del Consiglio Nazionale delle
Ricerche, Area della Ricerca di Pisa, Via Alfieri 1, San Cataldo,
56010, Ghezzano-Pisa, Italy \\
$^2$Center for Nonlinear Science, University of North Texas,
P.O. Box 311427, Denton, Texas 76203-1427\\
$^3$Istituto di Biofisica del Consiglio Nazionale delle
Ricerche, Area della Ricerca di Pisa, Via Alfieri 1, San Cataldo,
56010, Ghezzano-Pisa, Italy\\
$^4$Dipartimento di Fisica dell'Universit\`{a} di Pisa and
INFM, via Buonarroti 2, 56127 Pisa, Italy \\
$^5$International School for Advanced Studies SISSA-ISAS and INFM, 
Via Beirut 2-4, 34014 Trieste, Italy }


\begin{abstract}
We study an idealized version of intermittent process
 leading the fluctuations of a stochastic dichotomous variable $\xi$. 
It consists of an overdamped and symmetric potential well with 
a cusp-like minimum. The right-hand and left-hand portions of the potential
 corresponds to $\xi = W$ and $\xi = -W$, respectively. 
When the particle reaches this minimum is  injected back to a different
 and randomly chosen position, still within the potential well.
We build up the corresponding Frobenius-Perron equation and we evaluate
 the correlation function of the stochastic variable $\xi$, called
$\Phi_{\xi}(t)$. We assign to the potential well a form yielding 
$\Phi_{\xi}(t) = (T/(t + T))^{\beta}$, with $\beta > 0$. 
Thanks to the symmetry of the potential, there are no biases, and we 
limit ourselves to considering correlation functions with an even 
number of times, indicated for concision, by
$\langle 12 \rangle$, $\langle 1234 \rangle$ and, 
more, in general, by $\langle 1 ... 2n \rangle$. 
The adoption of a 
formal treatment, based on density, and thus of the operator driving 
the density time evolution, establishes a prescription for the 
evaluation of the correlation functions, yielding 
$\langle 1 ... 2n \rangle = \langle 1 2 \rangle ... \langle (2n-1) 2n \rangle$. 
We study the same dynamic problem using trajectories, 
and we establish that the resulting two-time correlation function 
coincides with that afforded by the density picture, as it should. We 
then study the four-times correlation function and we prove that in 
the non-Poisson case it departs from the density prescription, 
namely, from $\langle 1234 \rangle = \langle 12 \rangle \langle 34 \rangle$. 
We conclude that this is the main 
reason why the two pictures yield two different diffusion processes, 
as noticed in an earlier work [M. Bologna, P. Grigolini, B.J. West, 
Chem. Phys. {\bf 284}, (1-2) 115-128 (2002)].
\end{abstract}

\maketitle

\section{Introduction}

The assumption of the equivalence between trajectories and density 
is one of the major tenets of modern physics. In fact, it   is of 
fundamental importance in statistical mechanics, where the concept 
of density was originally
born, as a consequence of the Gibbs perspective resting on the use of
 infinitely many copies of the same dynamic system \cite{penrose}.
The concept of density, under the form of statistical density matrix 
  is also at the basis of the foundation of quantum mechanics and of decoherence
 theory \cite{zeh} that enables us to interpret all the statistical processes
 occurring in nature as being compatible with 
the unitary time evolution of the whole Universe. 
As already pointed out by Joos years ago \cite{joos}, the derivation 
of equation of motion for density, non diagonal in the position representation
of the density matrix, does not lead to the unique result of letting 
trajectories emerge. On one side, this seems to make unnecessary to
 alter the quantum-mechanical law of evolution so that 
localized states for macro-objects naturally emerge \cite{ghirardi} 
and, on the other side, generates the conviction that there is no more need
for the postulate of wave-function collapse \cite{tegmark}.

It comes to be, therefore,  a great surprise that the authors of 
Ref. \cite{maurobologna} found a conflict between trajectories and 
density picture to exist in a case of non-ordinary statistical mechanics. 
This has to do with the diffusion process generated by a fluctuating variable
$\xi(t)$, and consequently with the very simple equation of motion
\begin{equation}
\frac{dx}{dt} = \xi(t).
\label{verysimpleequationofmotion}
\end{equation}
The anomalous nature of the resulting diffusion process is generated
 by the fact that the correlation function of the stochastic variable is
\begin{equation}
\label{correlationfunction}
\Phi_{\xi}(t) =  \left(\frac{T}{t+T}\right)^{\beta},
\end{equation}
with
\begin{equation}
0 \leq \beta \leq 1.
\label{criticalcondition}
\end{equation}
Notice that the form  of Eq.(\ref{correlationfunction}) is essential for 
the realization of anomalous diffusion. For this form of anomalous diffusion
 to be of L\'{e}vy kind, it is also fundamental to assume the variable $\xi$
 to be dichotomous. In this paper, we assume, in fact, that 
$\xi$ can only have the values $W$ and $-W$. The L\'{e}vy nature of the
 resulting diffusion process can be established using either the Continuous
 Time Random Walk  \cite{zumofenandklafter1993} or, as pointed out in
 Ref. \cite{annunziato}, the Generalized Central Limit Theorem 
 \cite{gnedenkokolmogorov}. Both methods, however, are based on 
the direct use of trajectories, and consequently
are not yet a rigorous derivation of L\'{e}vy statistics from 
within the Liouville-like approach made necessary by the adoption of 
density picture  \cite{allegro}. This ambitious task was addressed 
by the authors of Ref. \cite{allegro}, but it became clear later  \cite{maurobologna} that the
generalized diffusion equation built up by these authors becomes 
compatible with L\'{e}vy statistics through the adoption
of a Markov approximation that forces it to depart significantly 
from the exact density time evolution. The form of this generalized diffusion equation is:
\begin{equation}
\label{generalizeddiffusionequation}
\frac{\partial}{\partial t} \sigma(x,t) = \langle\xi^{2}\rangle 
\int_{0}^{t} \Phi_{\xi}(t-t') \frac{\partial^{2}}{\partial x^{2}} \sigma(x,t') dt'.
\end{equation}

The main purpose of this paper is to identify the crucial 
property that is responsible for the conflict
between trajectories and density discovered 
by the authors of Ref. \cite{maurobologna}. 
However, to reach this important result in a satisfactory manner it 
is necessary to complete the program of the authors of Ref. \cite{maurobologna}. 
These authors proposed a Frobenius-Perron operator to describe the 
dynamics of the fluctuation process responsible for anomalous 
diffusion. We have to prove that this Frobenius-Perron operator 
yields the same correlation function as the trajectories produced by 
the same dynamics. This is usually taken for granted, but in this 
case, where the equivalence between the two pictures is questioned, 
it is of fundamental importance to prove that density and trajectory 
approach yield the same two-time correlation function. This is useful 
to identify the true source of breakdown. As already pointed out by 
the authors of Ref. \cite{mauro} (see also \cite{note}), once the Frobenius-Perron equation is 
established, we have to work with it without any more direct recourse 
to the concept of trajectories. Thus, we proceed as follows. We 
devote Section 2 to the derivation of the Frobenius-Perron operator. 
In Section 3 we derive the correlation function of 
Eq.(\ref{correlationfunction}) using 
trajectory. In Section 4 we derive the same correlation function from 
the Frobenius-Perron operator. As earlier stated, this is a crucial 
result, confirming that the source of discrepancy between trajectory 
and density has to be found somewhere else. In Section 5 we
derive Eq.(\ref{generalizeddiffusionequation}) and
in section 6 we identify 
the true source of the trajectories-density conflict. It is caused by 
the four-times correlation function, and plausibly, of higher order. 
We devote Section 7 to concluding remarks.

\section{An idealized model of intermittent dynamics and its Frobenius-Perron equation}

As done in Ref. \cite{maurobologna}, 
let us consider the following dynamical system. Let us consider
a variable $y$ moving within the interval $I = [0,2]$, within a sort of 
overdamped potential with a cusp-like minimum located at $y = 1$. 
If the initial condition of the particle is $y(0) > 1$, 
the particle moves from the right to the left towards the potential minimum. 
If the initial condition is $y(0) < 1$, then the motion of the particle 
towards the potential minimum takes place from the left to the right. 
When the particle reaches the potential bottom, it is injected back 
to an initial condition in the interval $I$. This initial condition, 
different from $y = 1$, is chosen in a random manner with uniform 
probability.
We thus realize a mixture of
 randomness and slow deterministic dynamics. The left and right portions of the
 potential $V(y)$ correspond to the laminar regions of turbulent dynamics, while
 randomness is concentrated at $y = 1$. In other words, this is an idealization
 of the map used by Zumofen and Klafter \cite{zumofenandklafter1993},
 which does not affect the long-time dynamics of the process, yielding only the benefit of
a neat distinction between random and deterministic dynamics.

The form of the Frobenius-Perron operator is the following
\begin{equation}
\label{evol_operator}
\frac{\partial}{\partial t}p(y, t) = - \frac{\partial}{\partial y}(\dot{y} p(y, t)) + C(t),
\end{equation} 
where, as we shall see,
\begin{equation}
C(t) \propto p(1,t),
\label{obviousformforinjectionback}
\end{equation}
which implies the random injection into any point $y$ of 
the interval $I$, with equal probability, due to the fact that $C(t)$ is independent of $y$. 
Eq.(\ref{evol_operator}) must fulfill the following physical conditions
\begin{equation}\label{define_ct}
\frac{d}{d t}\int_{I = [0, 2]}p(y, t) dy = 
\int_{I = [0, 2]}\frac{\partial}{\partial t}p(y, t) dy = 0.
\end{equation}
The condition (\ref{define_ct}) means that the evolution 
operator preserves the {\it mass} of the distribution
that without loss of generality is assumed to be given by
\begin{equation}
\int_{[0,2] }p(y,t)dy = 1.
\label{normalizationcondition}
\end{equation}
To ensure that plugging Eq.(\ref{evol_operator}) into Eq.(\ref{define_ct}) 
yields the natural condition of Eq.(\ref{obviousformforinjectionback}) and, 
at the same time, the probability conservation of Eq.(\ref{normalizationcondition}) 
\cite{note}, we are
 forced to adopt the following condition for $\dot{y}$,
\begin{equation}\label{new_doty}
\dot{y} = \lambda[\Theta(1-y)y^z - \Theta(y-1)(2-y)^z] 
+ \frac{\Delta_y(t)}{\tau_{random}}\delta(y-1).
\end{equation}
The function $\Theta(x)$ is the ordinary Heaviside step function, 
$\Delta_{y}(t)$ is a random
 function of time that can get any value
of the interval $[-1,+1]$, and $\tau_{random}$ is the injection time that must fulfill
 the condition of being infinitely smaller than
the time of sojourn in one of the laminar phases. 
Eq.(\ref{new_doty}), as well as Eq.(\ref{evol_operator}), 
is the combination of two processes, one deterministic and 
regular, the motion within the laminar phase, and the other, 
expressed by the second term on the right hand side of Eq.(\ref{new_doty}), 
totally random.

For the calculations that we shall make in Section 4, it is convenient to
 split the density $p(y,t)$ into symmetric and 
antisymmetric part with respect to $y=1$,
\begin{equation}
\label{anti-symm1}
p(y, t) = p_S(y, t) + p_A(y, t).
\end{equation}
Of course we always have that:
\begin{eqnarray} \label{anti-symm2}
\int_{[0, 2]}p_A(y, t)dy & = & 0 \nonumber\\ \int_{[0, 2]}p_S(y,t)dy & = & 1.
\end{eqnarray}
The Frobenius-Perron equation of Eq.(\ref{evol_operator}), with $\dot{y}$
 given by Eq.(\ref{new_doty}), expressed in terms of the symmetric and
 anti-symmetric part, yields the following two equations

\begin{eqnarray}\label{new_evol_symm}
\frac{\partial}{\partial t}p_S(y, t) & = & - \lambda
\Theta(1-y) \frac{\partial}{\partial y} [y^z p_S(y, t)] \nonumber \\
 & & + \lambda \Theta(y-1) \frac{\partial}{\partial y}[(2-y)^z p_S(y, t)]
+ C(t) \nonumber\\ 
\end{eqnarray}
and
\begin{eqnarray}\label{new_evol_anti}
\frac{\partial}{\partial t}p_A(y, t) & = & - \lambda
\Theta(1-y) \frac{\partial}{\partial y} [y^z p_A(y, t)] \nonumber \\
& & + \lambda \Theta(y-1) \frac{\partial}{\partial y}[(2-y)^z p_A(y, t)] .
\end{eqnarray}

Note that  Eq.(\ref{new_evol_symm}) yields as an  equilibrium state  the invariant distribution
$p_{0}(y)$,  whose explicit expression is easily obtained from Eq.(12) to be
\begin{equation}\label{inv_distr}
p_0(y) = \frac{2-z}{2}\left[\frac{\Theta(1-y)}{y^{z-1}} + \frac{\Theta(y-1)}{(2-y)^{z-1}}\right].
\end{equation}
The reader can easily check this to be the solution of Eq.(\ref{new_evol_symm}), by 
setting $C(t) =\lambda p_{0}(1)$, $p_S(y,t) = p_{0}(y)$, and using 
Eq.(\ref{inv_distr}) to determine $p_{0}(1)$ and $p_{0}(y)$.

  We denote this equilibrium state with the  quantum-like symbol
$|p_{0} \rangle$.  We refer ourselves to a notation where the corresponding
 left state, $\langle \tilde p_{0}|$ is nothing but the constant $1$. 
We write the correlation function $\Phi_{\xi}(t)$ under the quantum-like form
\begin{equation}
\label{property1}
W^{2} \Phi_{\xi}(t) = \langle\hat \xi(t) \hat \xi(0)\rangle = 
\langle\tilde p_{0}|\hat \xi \exp (\hat \Gamma t) \hat \xi|p_{0}\rangle,
\end{equation}
where $W^2 = \langle\hat \xi \hat \xi\rangle $. We note that when equilibrium is
 reached,  Eq.(\ref{new_evol_symm}) leaves room only for the state
 $|p_{0}\rangle$. All the other excited states would imply an out of 
equilibrium condition. Note that the diffusion process under study in this
 paper is stationary. This means that the correlation function is
 evaluated in the correspondence of the equilibrium condition $|p_{0}\rangle$. 
The consequence of this is significant. In fact the application of the
 operator $\hat \xi$  to $|p_{0}\rangle$ generates an antisymmetric state. 
The time evolution of this excited state 
is driven by the operator $\hat \Gamma$  defined by Eq.(\ref{new_evol_anti}). 
This means that the time evolution takes place in the antisymmetric space. 
When we apply to this time dependent state the operator $\hat \xi$ we
 obtain again a symmetric state. However, the only symmetric state 
available is $\langle\tilde p_{0}|$. This explains the form of Eq.(\ref{property1}).  
If we apply the same argument to the four-times correlation function we obtain,
 in the time ordered case $t_{1} \leq t_{2} \leq t_{3} \leq t_{4}$, 
\begin{equation}
\label{property2}
\langle\xi(t_{1})\xi(t_{2})\xi(t_{3})\xi(t_{4})\rangle = \langle\hat \xi(t_{2} - t_{1}) 
\hat \xi(0)\rangle\langle\hat \xi(t_{4} - t_{3}) \hat \xi(0)\rangle.
\end{equation}
This is the four-times case of a more general property that, 
using a concise notation, with an evident meaning, we write as
\begin{equation}
\langle 1 2 ....2n \rangle = \langle 1 2 \rangle ... \langle 2(n-1) 2n \rangle.
\label{evident}
\end{equation}
The authors of Ref.~\cite{maurobologna} proved that the property of
 Eq.(\ref{evident}) yields the generalized diffusion equation
 of Eq.(\ref{generalizeddiffusionequation}).
This confirms that Eq.(\ref{generalizeddiffusionequation}) 
is a legitimate and rigorous consequence 
of the Frobenius-Perron picture.

\section{The correlation function of the dichotomous fluctuation from the trajectory picture}

Let us focus our attention on Eq.(\ref{new_doty}) and consider
 the initial condition $y_0 \in [0, 1]$. Then, as it is 
straightforward to prove, the solution for $y < 1$ is:
\begin{equation}\label{solution}
y(t) = y_0\left(1-\lambda (z-1) y_0^{z-1} t\right)^{-1/(z-1)}.
\end{equation}
Using Eq.(\ref{solution}) and setting $y(T) = 1$, we find the
 time at which the trajectory reaches the point $y = 1$. The solution is:
\begin{equation}\label{arrive}
T = T(y_0) = \frac{1-y_0^{z-1}}{\lambda (z-1) y_0^{z-1}}.
\end{equation}
Note that the function  $\hat{\xi}(t)$ is determined by
$\hat{\xi}(t) = \hat{\xi}(y(t))$. Thus, from Eq.(\ref{arrive}) we obtain:
\begin{eqnarray}\label{xi_t}
\frac{\hat{\xi}(t)}{W} = [\Theta(1-y_0)\Theta(T(y_0) - t) \nonumber \\ 
-\Theta(y_0-1)\Theta(T(2-y_0) - t)] \nonumber\\
-\sum_{i = 0}^{+\infty} \mbox{sign}\left[\Delta_y\left(\sum_{k = 0}^i \tau_k \right)\right] 
\nonumber \\
\times \left[ \Theta\left(\sum_{k = 0}^{i+1}\tau_k -t\right) - 
\Theta\left(\sum_{k = 0}^i\tau_k -t\right) \right], \nonumber \\ 
\end{eqnarray}
where the times of sojourn in the laminar phases $\tau_i$'s read
\begin{eqnarray}\label{define_tau}
\tau_0 = T(y_0) =  \frac{1-y_0^{z-1}}{\lambda (z-1) y_0^{z-1}}\Theta(1-y_0) \nonumber \\
+ \frac{1-(2-y_0)^{z-1}}{\lambda (z-1) (2-y_0)^{z-1}}\Theta(y_0-1) \nonumber\\
\nonumber\\
\tau_{i \ge 1} =  \frac{1-[1+\Delta_y(\tau_i)]^{z-1}}{\lambda(z-1)[1+\Delta_y(\tau_i)]^{z-1}}
\Theta(-\Delta_y(\tau_i))  \nonumber\\
+ \frac{1-[1-\Delta_y(\tau_i)]^{z-1}}{\lambda(z-1)[1-\Delta_y(\tau_i)]^{z-1}}\Theta(\Delta_y(\tau_i)) .
 \nonumber\\ 
\end{eqnarray}
Then, we have:
\begin{eqnarray}\label{xit_xi0}
\frac{\langle \hat{\xi}(t) \hat{\xi}(0) \rangle}{W^2} =  
\langle [\Theta(1-y_0)\Theta(T(y_0) - t) \nonumber \\
+  \Theta(y_0-1)\Theta(T(2-y_0) - t)] \rangle + \nonumber\\
 \sum_{i = 0}^{+\infty} \left \langle \mbox{sign}(y_0-1) \mbox{sign}
\left[\Delta_y\left(\sum_{k = 0}^i \tau_k \right)\right] \right. \nonumber\\
\times \left. \left[ \Theta\left(\sum_{k = 0}^{i+1}\tau_k -t\right) - 
\Theta\left(\sum_{k = 0}^i\tau_k -t\right) \right] \right \rangle .
\end{eqnarray}
As pointed out in Section 2,  the calculation of the correlation function rests on averaging on the
 invariant distribution of Eq.(\ref{inv_distr}). 
As a consequence of this averaging the second term in(\ref{xit_xi0}) vanishes.  In fact, 
the quantity to average is antisymmetric, whereas the statistical weight is symmetric.

It is possible to write the surviving term as:
\begin{eqnarray}\label{xit_xi0_bis}
\frac{\langle \hat{\xi}(t) \hat{\xi}(0) \rangle}{W^2} & = & 
(2-z)\int_0^1\Theta\left(\frac{1-y^{z-1}}{\lambda(z-1)y^{z-1}} - t\right)\frac{1}{y^{z-1}}dy \nonumber\\
& = & (2-z)\int_0^{(1+\lambda(z-1)t)^{-1/(z-1)}} y^{-z+1} dy \nonumber\\
& = & (1+\lambda(z-1)t)^{-(2-z)/(z-1)} \nonumber\\
& \equiv & (1+\lambda(z-1)t)^{-\beta},
\end{eqnarray}
with
\begin{equation}\label{beta}
\beta = \frac{2-z}{z-1} .
\end{equation}
Since we focus our attention on $0 < \beta < 1$, 
we have to consider $3/2 < z < 2$. Note that the region $1 < z< 3/2$
does not produce evident signs of deviation from ordinary statistics. 
However, as we shall see in Section 5,
 an exact agreement between density and trajectory is recovered only at $z = 1$, when
the correlation function becomes identical to the exponential function $\exp(- \lambda t)$.
 Note also that Eq.(\ref{xit_xi0_bis}) becomes identical to Eq.(\ref{correlationfunction}) 
after setting the condition
\begin{equation}
\lambda (z-1) = \frac{1}{T}.
\label{constraint}
\end{equation}

\section{The correlation function of the dichotomous fluctuation from the density picture}

The result of the preceding section is reassuring, since it proves that 
the intermittent model we are using generates the wanted inverse power law 
form for the correlation function of the dichotomous variable $\xi(t)$. 
However, it is based on the adoption of trajectories.
In this Section we get an even more important result, 
this being the fact that the same correlation function
is derived from the use of the Frobenius-Perron 
equation of Eq.(\ref{evol_operator}), with no direct use of trajectories. 

To fix the ideas, let us consider the following system: a particle in the interval 
$[0, 1]$ moves towards $y=1$ following the prescription 
$\dot{y} = \lambda y^z$ and when it reaches $y = 1$ it is injected 
backwards at a random position in the interval. 
The evolution equation obeyed by the densities defined on this interval 
is the same as Eq.(\ref{evol_operator}), with $C(t) = \lambda p(1, t)$. 
This dynamic problem was already addressed in Refs. \cite{mauro,massi}, 
and solved using the method of characteristics detailed in Ref. \cite{goldenfeld}. 
Let us remind the reader that the solution afforded by 
the method of characteristics, in this case yields:
\begin{eqnarray}
\label{first_solution}
p(y, t) = \int_0^t\frac{p(1, \xi)}{g_z((t-\xi)y^{z-1})}d\xi \nonumber \\
+p \left(\left[1-\frac{y}{g_z(y^{z-1}t)}\right], 0\right) 
 \frac{1}{g_z(y^{z-1}t)},
\end{eqnarray}
where
\begin{equation}
g_z(x) \equiv (1+\lambda(z-1)x)^{z/(z-1)}.
\end{equation}

To find the quantity $\langle \hat{\xi}(t) \hat{\xi}(0)\rangle$ using only 
density, we have to solve Eqs.(\ref{new_evol_symm}) and (\ref{new_evol_anti}), 
which are equations of the same form as that yielding Eq.(\ref{first_solution}). 
For this reason we adopt the method of characteristics again. 
To do the calculation in this case, it  is convenient to adopt a frame 
symmetric respect to $y = 1$. Then, let us define $Y = y-1$. 
Using the new variable and Eq.(\ref{first_solution}), we find 
for Eqs.(\ref{new_evol_symm}) and (\ref{new_evol_anti}) the following solution

\begin{eqnarray}\label{solution1}
p_S(Y, t) & = &\int_0^t\frac{p_S(0, \xi)}{g_z((t-\xi)(1-|Y|)^{z-1})}d\xi + \nonumber\\ 
& & p_S \left(\left[1-\frac{1-|Y|}{g_z((1-|Y|)^{z-1} t)}\right], 
0\right) \times \nonumber\\ 
& & \times \frac{1}{g_z((1-|Y|)^{z-1} t)}
\end{eqnarray}
and
\begin{eqnarray}\label{solution2}
p_A(Y, t) & = & p_A \left(\mbox{sign}(Y)\left[1-
\frac{1-|Y|}{g_z((1-|Y|^{z-1}t))}\right],0\right) 
\times \nonumber\\ 
& & \times
\frac{1}{g_z((1-|Y|)^{z-1}t)}
\end{eqnarray}
Then, the solution consists of two terms:
\begin{itemize}
\item the former is an {\it even} term and 
is responsible for the long-time limit of the distribution evolution;
\item the latter is an {\it odd} term and it disappears in the
long-time limit. Let us point out why this feature is beneficial: 
Eq.(\ref{new_evol_anti}) does not contain the
injection term $C(t)$ and the equilibrium distribution function
 is given by the density distribution of Eq.(\ref{inv_distr}),
which is an {\it even} function, no
matter what the symmetry of the initial distribution function is.
\end{itemize}

As pointed out in Sections 2 (see Eq.(\ref{property1}) 
and Eq.(\ref{property2})), the 
correlation function of the operator $\hat{\xi}$ is determined by the 
time independent operator driving the time evolution of the 
anti-symmetric distribution. As proved in Section 5, this property is 
of fundamental importance for a rigorous
derivation of the generalized diffusion equation of 
Eq.(\ref{generalizeddiffusionequation}). In fact, 
this property means that the anti-symmetric space is the irrelevant 
space, to be traced out for the derivation of this important 
diffusion equation. The direct calculation of this correlation 
function yields:
\begin{eqnarray}\label{correlation}
\langle \hat{\xi}(t) \hat{\xi}(0) \rangle & = & (2-z) \left. 
\int_0^1\frac{1}{(1-\xi)^{z-1}} \right|_{\xi = 1-\frac{1-Y}{g_z((1-Y)^{z-1}t)}} 
\times \nonumber\\ 
& & \times \frac{1}{g_z((1-Y)^{z-1}t)}dY
\end{eqnarray}
The integral (\ref{correlation}) is exactly solvable and leads to:
\begin{equation}\label{correl}
\langle \hat{\xi}(t) \hat{\xi}(0) \rangle = (1+\lambda (z-1)t)^{-(2-z)/(z-1)}
\end{equation}
which is the same result as that found in Section 3, using the trajectories.
In a similar fashion, it is possible to calculate $\langle Y(t)Y(0)\rangle$: 
its temporal behavior is a power law with the same 
exponent as that of Eq.(\ref{correl}).

\section{The generalized diffusion equation revisited}

In this section we shall derive again the generalized diffusion equation
 of Eq.(\ref{generalizeddiffusionequation}) in a way that is more closely 
connected with the formalism of Sections  2 to 4.  We define
\begin{equation}
|p_{1}\rangle = \frac{\hat \xi}{W} |p_{0}\rangle.
\label{antisymmetricstate}
\end{equation}
This is an antisymmetric state. Consequently its time evolution is driven 
by the operator $\hat \Gamma$ defined by Eq.(\ref{new_evol_anti}). 
We now assume \cite{driebe} that the operator $\hat \Gamma$ has a complete basis set of 
eigenvectors with a non-necessarily discrete spectrum, even if adopt for simplicity a 
discrete notation, setting aside the easy problem of moving from the discrete to the 
continuous condition. Let us assume
\begin{equation}
\label{righteingenstate}
\hat \Gamma|\mu\rangle = \gamma_{\mu}|\mu\rangle 
\end{equation}
and
\begin{equation}
\label{lefteigenstate}
\langle {\tilde \mu} | \hat\Gamma^{+} = \langle {\tilde \mu} | \gamma_{\mu}^{*}.
\end{equation}
so that the biorthonormal condition
\begin{equation}
\label{biorthonormal}
\langle \tilde \mu | \mu' \rangle = \delta_{\mu, \mu'}
\end{equation}
applies.
All this allows us to write the correlation 
function $\Phi_{\xi}(t)$ under the following attractive form
\begin{equation}
\label{crucialexpresio}
\Phi_{\xi}(t) = \langle \tilde p_{1}| \exp (\hat \Gamma t) |p_{1} 
\rangle =  \sum_{\mu} \langle \tilde p_{1} | \mu \rangle  \langle 
\tilde \mu|p_{1} \rangle \exp(\gamma_{\mu} t)   
\end{equation}

We are now in a position to derive the generalized diffusion equation of 
Eq.(\ref{generalizeddiffusionequation}) from a Liouville-like equation. 
It is evident that the Liouville-like equation of the whole system, corresponding 
to the Frobenius-Perron treatment of the system of variables $\xi$ and $y$ 
of the preceding sections, is given by
\begin{equation}
\frac{\partial}{\partial t} \Pi(x,\xi,y,t) = 
\left ( -\xi \frac{\partial}{\partial x} + \Gamma \right ) \Pi(x,\xi,y, t).
\label{totalequation}
\end{equation}
Let us adopt the quantum-like formalism and let us set
\begin{equation}
\rho_{\mu} \equiv \langle \tilde \mu |\Pi(t) \rangle
\end{equation}
and
\begin{equation}
\sigma (x,t) \equiv \rho_{0} (x,t) \equiv \langle {\tilde p}_{0} |\Pi(t) \rangle.
\end{equation}
Thus, by multiplying Eq.(\ref{totalequation}) by $\langle {\tilde p}_{0}|$ 
and $\langle \tilde \mu|$,we get
\begin{equation}
\label{first}
\frac{\partial}{\partial t} \sigma (x,t) = -W \sum_{\mu} \langle{\tilde p}_1 | 
\mu \rangle \frac{\partial}{\partial x} \rho_{\mu}(x,t) 
\end{equation}
and
\begin{equation}
\label{second}
\frac{\partial}{\partial t} \rho_{\mu}(x,t)  = -W \frac{\partial}{\partial x} 
\sigma (x,t) + \gamma_{\mu} \rho_{\mu}(x,t),
\end{equation}
respectively. 
The solution of Eq.(\ref{second}) is
\begin{equation}
\label{solution_bis}
\rho_{\mu}(x,t) = -  W \int_{0}^{t} \exp(\gamma_{\mu} (t - t') )
\frac{\partial}{\partial x} \sigma(x,t') dt'.
\end{equation}
By plugging Eq.(\ref{solution_bis}) into Eq.(\ref{first}) we derive the
 fundamental equation of Eq.(\ref{generalizeddiffusionequation}).
 It is easy to prove \cite{maurobologna} that this equation yields a
 hierarchy of moments $\langle x^{2n}(t)\rangle$, with $n = 1, 2, ..., \infty$,
 which coincides with the hierarchy that would be generated by a fluctuation
 $\xi(t)$ with the correlation functions obeying the prescription of Eq.(\ref{evident}). 
 In the next section
 we shall see that this conflicts with the prescriptions generated by 
the adoption of trajectories. The arguments of Section 6 are limited to the case
of four-times correlation function. However, we think that this is enough to 
identify the source of the conflict between trajectories and density.

\section{four-times correlation function}

In this section we show that the four-times correlation function
stemming from probabilistic arguments and trajectories do not fulfill
Eq.(\ref{property2}). We shall see that we can write the
four-times correlation function in terms of a different combination
of two-times correlation function. In this way we can connect
higher-order correlation function to the waiting time distribution
$\psi(t)$ previously defined. Note that the waiting time
distribution and the correlation function are linked via \cite{Geisel}
\begin{equation}\label{geisel}
\Phi_{\xi}(t_2 -t_1)=\frac{1}{\langle T \rangle}\int\limits_{t_2 -t_1}^{\infty} 
[T-(t_2 -t_1)]\psi(T) dT .
\end{equation}
This means that the correlation function can be viewed
as the probability of finding $t_1$
and $t_2$ in the same laminar region.  
In fact, the occurrence of one or more transitions, between $t_{1}$ 
and $t_{2}$, would make the correlation function vanish due to the 
averaging on uncorrelated fluctuations, either positive or negative.
Using similar arguments
the four-times correlation function can be viewed  as the probability
that both $t_1$ and $t_2$ belong to one laminar region and 
so do $t_3$ and $t_4$. In the following
we shall use the notation $p(ij)$ as the probability that $t_i$ and
$t_j$ belong to the same laminar region.
We shall also use the notation $p(\bar{ij})\equiv 1-p(ij)$ as the probability
that at least one transition occurs between $t_i$ and $t_j$. 
With the usual Bayesian notation we indicate with $p(A,B)$ the
joint probability of events $A$ and $B$ and with $p(A|B)$ the
conditional probability of $A$ given $B$ (thus implying $p(A,B)\equiv p(B) p(A|B)$).
It is easy to cast the four-times correlation function in our notation as (with
$W^2=1$):
\begin{equation}
\label{4timestrajectories}
\langle  \xi(t_1) \xi(t_2) \xi(t_3) \xi(t_4) \rangle=
p(12,34)
\end{equation}
\[
= p(14) + p(\bar{23}) p(12|\bar{23})) p(34|\bar{23})
\]
\[
=p(14)+\frac{(p(12)-p(12,23))(p(34)-p(34,23))}{1-p(23)}
\]
\[
=\Phi_{\xi}(t_4-t_1)
\]
\[
 + \frac{ ( \Phi_{\xi}(t_2-t_1) - \Phi_{\xi}(t_3-t_1) ) (\Phi_{\xi}(t_4-t_3) 
- \Phi_{\xi}(t_4-t_2) ) }
{1-\Phi_{\xi}(t_3-t_2)}, 
\]
where we used the general property:
\begin{equation}
p(A|\bar B) = \frac{p(A)-p(A,B)}{1-p(B)}
\end{equation}
and the fact that in our problem
\begin{equation}
p(ij,jk)= p(ik).
\end{equation}

The leading order of Eq.(\ref{4timestrajectories}) is the
first term, namely $\Phi_{\xi}(t_4-t_1)$. This yields, for the
fourth moment
\begin{equation}
\label{4thmoment}
\langle x^4 (t) \rangle = 8 \int \limits_{0}^{t} dt_4 
\int \limits_{0}^{t_4} dt_3 \int \limits_{0}^{t_3} dt_2 
\int \limits_{0}^{t_2} dt_1 \langle  \xi(t_1) \xi(t_2) \xi(t_3) \xi(t_4) \rangle,
\end{equation}
and, if the correlation function has a slow decay as 
in (\ref{correlationfunction}),  $\langle x^4 (t) \rangle \propto t^{4-\beta}$.
Notice that if condition (\ref{property2}) held true, we
would have  $\langle x^4 (t) \rangle \propto t^{4-2\beta}$.
Generalizing the 4-times treatment to the $2n$-times, the trajectory 
prescription yields
\begin{equation} 
\langle x^{2n} (t) \rangle \propto t^{2n-\beta},
\end{equation}
as recently shown in Ref. \cite{aging}. On the other hand, the
factorization of the even correlation functions, which is exact in the
density treatment, leads to $\langle x^{2n} (t) \rangle \propto t^{2n(1-\beta/2)}$,
fulfilling (\ref{generalizeddiffusionequation}).
This latter property reflects the fact that the solution
of the generalized diffusion equation (\ref{generalizeddiffusionequation})
has a solution which rescales with time, while its trajectory counterpart
does not obey an exact rescaling \cite{aging}.

Remarkably, in the exponential case, with some tedious algebra it is 
straightforward to show that 
\begin{equation}\label{factorized4-2}
\langle  \xi(t_1) \xi(t_2) \xi(t_3) \xi(t_4) \rangle
= \langle  \xi(t_1) \xi(t_2) \rangle \langle \xi(t_3) \xi(t_4) \rangle,
\end{equation}
which coincides with prescription (\ref{property2}).
With boring but straightforward arguments it is possible to prove 
that Eq.(\ref{factorized4-2}) 
yields the more general condition of Eq.(\ref{evident}). In 
conclusion, in the exponential case the agreement  between density 
and trajectories is complete.

\section{Concluding Remarks}

The discrepancy between trajectories and density is a 
surprising result. 
The first report of this disconcerting property was given in
Ref. \cite{maurobologna}.
However, in that paper the derivation of the generalized 
diffusion equation, Eq.(\ref{generalizeddiffusionequation}),
 from a Frobenius-Perron picture was not as accurate as in 
the present paper, thereby leaving room to the criticism that this
 equation is not compatible with a rigorous Frobenius-Perron picture. 
This is the reason why Sections 2, 3 and 4 have been devoted to a careful
 discussion of the Frobenius-Perron picture. Of remarkable interest is the content of
 Section 4 which derives the crucial correlation function of Eq.(\ref{correlationfunction})
 without using the concept of trajectory. There is no doubt, therefore, that the density
 picture adopted is the proper representation of the process under study.

 The importance of the Gibbs picture for statistical 
mechanics is well known, and it is well known that the ergodic 
properties have been introduced to replace the average over infinitely 
many copies of the same system with a more realistic average in time 
\cite{ma}. However, there is no guarantee that the two pictures are 
equivalent, in general. Thus, to a first sight the reader might think 
that the breakdown of the equivalence between trajectories and 
density rests on the failure of the ergodic condition, a relatively 
trivial property. The effect discovered by the authors of Ref. \cite{maurobologna} 
does not have anything to do with the breakdown of the ergodic 
condition. If we focus our attention on the variable $y$ and on the 
Frobenius-Perron picture of Section 2, we do not find any 
disagreement between trajectories and density,  the results of 
Sections  3 and 4 lead us to expect that the numerical study of a 
bunch of trajectory yields for the time evolution of a non equilibrium 
probability distribution the same time evolution as that prescribed 
by the Frobenius-Perron operator . On top of that the variable $y$ is 
ergodic and the generalized diffusion equation of 
Eq. (\ref{generalizeddiffusionequation}) is based on 
the correct assumption that the Gibbs averages are equal to the time 
average on time of a single trajectory.  This is a correct 
assumption.  The reason of the failure of 
Eq. (\ref{generalizeddiffusionequation}) to recover the 
results of numerical calculations based on trajectories is much 
deeper and subtler than the failure of the ergodic condition. This 
reason has to do with the fact that the higher-order correlation 
functions are expressed in terms of the two-time correlation 
functions in two distinct ways, the trajectory and the density way. 
We have shown in details that the four-times correlation function 
compatible with the generalized diffusion equation of 
Eq. (\ref{generalizeddiffusionequation}) is the 
product of two second-order (two-times) correlation functions. In the 
non-exponential case, the four-times correlation function generated 
by the trajectories breaks this condition.
Extending the kind of arguments used in Section 6 to correlation 
functions of higher order involves much more complicated 
calculations, but the ballistic contribution always shows up, and the 
formulas can be expressed in terms of pairs of two-time correlation 
functions, although by means of expressions of increasing length. In 
conclusion, even if for simplicity it is not done here, it is 
possible to prove that in the case of exponential relaxation, the 
factorization condition is fulfilled by the correlation functions of 
any order. This means that the exponential case is the only one where 
the equivalence between trajectory and density is complete.

What might be the consequence of this result? In the introduction we
 mentioned the possible consequences on the foundation of quantum mechanics.
 One might be tempted to conclude that the foundation of L\'{e}vy processes
 on the basis of classical trajectories rules out the possibility of
 deriving L\'{e}vy diffusion from a quantum mechanical treatment, in
 conflict with the conclusions of several papers that suggest the
 opposite to be true \cite{opposite1,opposite2,opposite3,opposite4}.
 We have to point out that all these papers studied the kicked rotor,
 and with this dynamical system the condition of ballistic motion
 for extended times is realized by a trajectory sticking to the
 border between chaotic sea and acceleration islands. 
This sticking is actually due to the fact that the classical
 trajectories keeps exploring fractal regions of decreasing size. 
The arguments used to justify the birth of anomalous diffusion 
rest on classical trajectories, and the quantum breakdown of
 this condition of anomalous diffusion is due to the broadening
 of the quantum wave function that prevents it from exploring
 border regions of even infinitesimally small size. 
It is plausible that the realization of L\'{e}vy statistics 
cannot be complete, and that the breakdown of anomalous 
diffusion takes place in the very moment when tunneling 
processes are activated. All this seems to prove that the 
papers of Refs. \cite{opposite1,opposite2,opposite3,opposite4} 
cannot be used to argue against the conclusion of this paper, 
and cannot be used to support them either. 

We think that the conclusions of this paper might be of 
interest for quantum mechanics at a different level. 
The experiment on resonant fluorescence on single atoms 
have revealed experimentally the occurrence of quantum 
jumps \cite{knight}. Nevertheless the discovery that the 
Bloch equations consists of two contributions, one 
compatible with the unitary nature of quantum mechanics, 
and one equivalent to the von Neumann measurement 
postulate, makes it plausible to reach the conclusion 
that, as a relevant example, the reader can find in the 
introduction of the book of Ref. \cite{weiss}. 
This is that the hypothesis of spontaneous collapses is unnecessary. 
The contraction on the irrelevant degrees of freedom would be enough 
to mimic properties that are naively interpreted as a manifestation
 of collapses.

This view is generally accepted, but it is challenged by new 
discoveries. We note that the unraveling process compatible with the 
occurrence of quantum jumps rests its foundation on the connection 
between random walker trajectories and master equation, established 
years ago by the authors of Ref. \cite{bedeaux}. These authors proved that an 
exponential waiting time distribution implies a Markovian master 
equation. Using the more modern jargon of de-coherence literature 
\cite{knight} we can say that the exponential waiting time distribution is 
compatible with the quantum mechanical master equations that have the 
Lindblad structure. However, recent experimental studies on 
semiconductor quantum dots \cite{barkai,note2} reveal intermittent signals 
corresponding to a bright state emitting many photons and to a dark 
state with no photon emission. The waiting time distributions of 
these two states turn out to exhibit inverse power law properties for 
several decades. There are already conjectures that this strong 
deviation from an exponential behavior might be due to the existence 
of slow modulation of barrier widths or heights \cite{kuno}. 

It is worth 
noticing that the authors of Ref. \cite{bolognaCSF} used this modulation 
hypothesis to discuss an ideal physical problem that it is very 
similar to the blinking semiconductors. These authors studied the 
case where a given Pauli matrix $\sigma_{x}$  is forced by the 
environment to yield a symbolic sequence of numbers, corresponding to 
its two eigenvalues, a sequence that can be thought of as 
corresponding to that of a blinking semiconductor, with, for 
instance,  the eigenvalue $1$ standing for light, and the eigenvalue 
$-1$  for darkness. This fluctuating dipole drives the motion of 
another 1/2 spin system, serving a detection purpose. 
Actually, this 
second spin system is essentially equivalent to the Kubo stochastic 
dipole \cite{kubo} used by the authors of Ref. \cite{barkai}. 

The authors of Ref. \cite{bolognaCSF} 
found a surprising effect. They studied the relaxation properties of 
the detecting dipole driven by the intermittent signal, with a modulation
assumption about the origin of its non-Poisson nature that fits the 
conjectures made to explain the properties of blinking quantum dots 
\cite{kuno}. The waiting time distribution was assumed to have a diverging 
second moment, a property shared indeed by the blinking quantum dots 
\cite{barkai}. The calculation of these spectroscopy properties was done in 
two distinct ways. The former was compatible with the density 
prescription, with the only constraint that the fluctuation 
correlation function has to coincide with that of the symbolic 
sequence realized by the environmental induced process. The latter 
calculation was done, on the contrary, noticing that the spectroscopy 
property can also be expressed through a characteristic function. The 
explicit analytical form of this characteristic function is 
determined by the central limit theorem, either normal, in the case 
of ordinary statistical mechanics, or generalized, in the case of 
interest, that is, the case of waiting time distributions with a 
diverging second moment.  This latter way of doing the calculation 
yielded a result conflicting with the former, with the detecting 
dipole correlation function decaying as an exponential rather than as 
an inverse power law. We think that the present article affords a 
satisfactory explanation of why the two methods adopted yield 
different results. This is so because only the exponential relaxation 
make the correlation function stemming from the density picture 
identical to those produced by trajectories. A deviation from the 
Poissonian condition seems to have dramatic consequences. We hope 
that this paper might serve the purpose of triggering further 
investigation on this important issue.

\section{acknowledgments} 
P.G. 
thankfully acknowledges
 financial support from ARO, through Grant DAAD19-02-0037.


\begin{thebibliography}{99}

\bibitem{penrose} O. Penrose, Rep. Prog. Phys. {\bf42}, 1938 (1979)

\bibitem{zeh} D. Giulini, E. Joos, C. Kiefer, J. Kupsch, I.-O.Stamatescu, 
and H.D. Zeh, \emph{Decoherence and the Appearance of a Classical World in Quantum Theory}, 
Springer, Berlin (1996). 

\bibitem{joos} E. Joos, Phys. Rev. D {\bf 36}, 3285 (1987).

\bibitem{ghirardi} G.C. Ghirardi, A. Rimini, and T. Weber, Phys. Rev. D {\bf 34}, 470 (1986). 

\bibitem{tegmark} M. Tegmark, J. A. Wheeler, arXiv:quant-ph/0101077. 

\bibitem{maurobologna} M. Bologna, P. Grigolini, B. J. West, Chem. Phys. {\bf 284}, 115 (2002). 

\bibitem{zumofenandklafter1993} G. Zumofen and J. Klafter, Phys. Rev. E {\bf 47}, 851 (1993). 

\bibitem{annunziato} M. Annunziato, P. Grigolini, Phys. Lett. A {\bf 269}, 31 (2000). 

\bibitem{gnedenkokolmogorov} B. V. Gnedenko, A. N. Kolmogorov,
\emph{Limit Distributions for Sums of Indipendent Random Variables}, 
Addison-Wesley Publisher, Cambridge (1954). 

\bibitem{allegro} P. Allegrini, P. Grigolini, B. J. West, Phys. Rev. E {\bf 54}, 4760 (1996). 

\bibitem{mauro} M. Bologna, P. Grigolini, M. Karagiorgis and A. Rosa, 
Phys. Rev. E {\bf 64}, 016223 (1-9) (2001).

\bibitem{note} It is important to stress that in Ref. \cite{maurobologna} 
the definition of the Frobenius-Perron operator was fraught with some inconsistencies
concerning the back injection term, generating a conflict  
with the mass conservation constraint of 
 Eq.(\ref{normalizationcondition}). This did not have any consequence on 
the main result reached by these authors, since their demonstration 
rests on  the assumption that a Frobenius-Perron operator with 
the properties of Eq.(\ref{evident}) exists. This paper proves that this 
assumption is correct, and determines the correct  
form of the Frobenius-Perron operator as well. 


\bibitem{massi} M. Ignaccolo, P. Grigolini, A. Rosa, Phys. Rev. E {\bf 64}, 026210 (1-11) (2001) .

\bibitem{goldenfeld} N. Goldenfeld, 
\emph{Lectures on Phase Transitions and the Renormalization Group}, 
Addison-Wesley Publisher, Cambridge (1992).

\bibitem{driebe}  D. J. Driebe, \emph{Fully Chaotic Maps and Broken Time Symmetry}, 
Kluwer Academic Publishers, Dordrecht ( 1999).


\bibitem{Geisel} T. Geisel, J. Nierwetberg, and A. Zacherl, Phys. Rev. Lett. {\bf 54}, 616 (1985).

\bibitem{aging} P. Allegrini, J. Bellazzini, G. Bramanti, M. Ignaccolo, P. Grigolini, and J. Yang, Phys Rev. {\bf 66}, 015101 (R) (2002). 

\bibitem{ma} S.-K, Ma, \emph{Statistical Mechanics}, World Scientific Publishing, Singapore (1985).

\bibitem{opposite1} R. Roncaglia, L. Bonci, B.J. West, P. Grigolini,  Phys. Rev. E {\bf 51}, 5524 (1995).

\bibitem{opposite2} L. Bonci, P. Grigolini, A. Laux, R. Roncaglia, Phys. Rev. {\bf 59}, 7231 (1996). 

\bibitem{opposite3} M. Stefancich, P. Allegrini, L. Bonci, P. Grigolini, and B.J. West,  Phys. Rev. E {\bf 57}, 6625 (1998).

\bibitem{opposite4} A. Iomin, G.M. Zaslavsky, Chem. Phys. {\bf 284}, 3 (2002). 

\bibitem{knight} M.B. Plenio and P. L. Knight, Rev. Mod. Phys. {\bf 70}, 101 (1998). 

\bibitem{weiss} U. Weiss, \emph{Quantum Dissipative Systems}  World Scientific, Singapore (1999). 

\bibitem{bedeaux} D. Bedeaux, K. Lakatos-Lindenberg, and K. E. Shuler, J. Math. 
Phys. {\bf12}, 2116 (1971).

\bibitem{barkai} Y. Jung, E. Barkai, R. J. Silbey, Chem. Phys. {\bf 284}, 181 (2002);


\bibitem{note2} The experimental literature on the subject of fluorescence 
intermittency phenomena is rapidly growing and we refer the reader to 
the excellent work of Ref. \cite{barkai} for experimental references as well 
as for a theoretical treatment of this challenging problem.

\bibitem{kuno} M. Kuno, D.P. Fromm, H.F. Hamann, A. Gallagher, 
and D.J. Nesbitt, J. Chem. Phys. {\bf 115}, 1028 (2001).

\bibitem{bolognaCSF} M. Bologna, P. Grigolini, L. Palatella, Marco Pala, 
"Decoherence, wave function collapses and non-ordinary statistical 
mechanics", to appear in Chaos, Solitons and Fractals, publication top priority 
(2002).

\bibitem{kubo} R. Kubo, J. Phys. Soc. Jpn. {\bf 9}, 316 (1954).


\end{thebibliography}
\end{document}